\documentclass[twocolumn,showpacs,preprintnumbers,amsmath,amssymb]{revtex4}

\usepackage{graphicx}
\usepackage{dcolumn}
\usepackage{bm}

\begin{document}

\def \be {\begin{equation}}
\def \ba {\begin{eqnarray}}
\def \ee {\end{equation}}
\def \ea {\end{eqnarray}}
\def \lsm {L$\sigma$M}
\def\bea{\begin{eqnarray}}
\def\eea{\end{eqnarray}}
\def\UA1{$U_A(1)$}\def\U3U3{$U(3)_L\times U(3)_R$}\def\SU3SU3{$SU(3)_L\times SU(3)_R$}
\def \S {$\Sigma$}
\def\lsim{\;\raise0.3ex\hbox{$<$\kern-0.75em\raise-1.1ex\hbox{$\sim$}}\;}
\def\gsim{\raise0.3ex\hbox{$>$\kern-0.75em\raise-1.1ex\hbox{$\sim$}}}

\author{Nils A. T\"ornqvist \\
Department of Physical Sciences, \\ University of Helsinki, POB
64, FIN--00014}
\title{The Scalar Mesons,  Symmetry Breaking, Three Colors and Confinement}
\begin{abstract}
The same, well known, $\det\Sigma+\det\Sigma^\dagger$ term in
effective theories, which 't Hooft showed is generated by
instantons in QCD and which resolves the \UA1 problem giving mass,
in particular to the $\eta'$ is for three light flavors shown to
give three classical minima along the $U_A(1)$ circle. The three
minima are related to the center $Z(3)$ of SU(3). The term also
contributes, in a similar way as the diquark model of Jaffe, to an
inverted scalar mass spectrum for the light scalars. The three
vacua suggests a connection to the strong $CP$ problem and
confinement.
\end{abstract}
\maketitle

It is widely believed that QCD with three nearly massless light
quark flavors  explain the well-known approximate \SU3SU3 chiral
symmetry seen in the light meson mass spectrum. Our  present
understanding of the  symmetry breaking involves three basic
mechanisms:

\begin{itemize}
\item[(i)] Spontaneous symmetry breaking in the QCD vacuum, which
gives rise to a near flavor symmetric $<\bar qq>$ condensate and
an octet of (would-be massless) Goldstone pseudoscalars.

\item[(ii)] A contribution from the gluon anomaly,
which explicitly breaks the axial symmetry $U_A(1)$ in \U3U3, and
which gives in, particular, mass to the $\eta'$\cite{Hooft}.

\item[(iii)] Small chiral quark masses $m_u, m_d, m_s$
from the electro-weak sector, which give the pseudoscalar octet
states a small mass and break flavor symmetry. A large $m_s/m_d$
mass  ratio together with the anomaly term (ii) also splits the
$\eta$ from the pion, which saves isospin symmetry in spite of the
large $m_d/m_u$ chiral quark mass ratio.
\end{itemize}

In effective theories for scalar and pseudoscalar mesons one
models the global \U3U3\ symmetry by potential terms. Including up
to dimension four terms one writes \footnote {We recall that by
putting the pseudoscalars ($p$) into the anti-Hermitian part of
$\Sigma=s+ip$ the $\gamma_5$ in $\bar u (s+i\gamma_5p) u$
disappears since we can write it as $\bar u [\frac 1
2(1-\gamma_5)\Sigma + \frac 1 2(1+\gamma_5)\Sigma^\dagger] u
$=$\bar u_L\Sigma u_R+\bar u_R\Sigma^\dagger u_L$. The parity
transformation of $\gamma_5$ is thus just complex conjugation,
$\Sigma \to \Sigma^* $, while $CP$ is represented by $\Sigma \to
\Sigma^\dagger$. From this it is also clear that $\Sigma$
transforms as $\Sigma \to U_L\Sigma U_R$, and $\Sigma^\dagger \to
U_R\Sigma^\dagger U_L$ under $U(3)_L\times\ U(3)_R$, from which
the invariance of the potential (\ref{U3U3}) follows.}
\begin{equation}\label{U3U3}
 { V_{\it{U3U3}}} =
 \frac {\mu^2} 2{\rm Tr} [\Sigma\Sigma^\dagger]
 +\lambda {\rm Tr[\Sigma\Sigma^\dagger\Sigma\Sigma^\dagger]} +\lambda' ({\rm
Tr}[\Sigma\Sigma^\dagger])^2, \end{equation} where $\Sigma$ is the
usual $3\times 3$ matrix containing the scalar ($s$) and
pseudoscalar ($p$) nonets. (Denoting the nonet members by $s_k$
and $p_k $ for $ k= 0$ to $ 8,$ one has $ \Sigma =  \sum_k
(s_k+ip_k)\lambda_k$, where $\lambda_k$ are Gell-Mann matrices).

If $\mu^2$ has the "wrong sign" $\mu^2<0$  eq.(1) predicts the
often quoted spontaneous symmetry breaking with a nonet of
massless pseudoscalars. But, in this case the $U(1)$
problem\cite{WeinU1} arises. There is "too much symmetry", the
axial $U_A(1)$ problem appears and  the $\eta$ and $\eta '$ become
massless.

To have a realistic zeroth order \SU3SU3\ model, one must follow
the step (ii) above and break the axial $U_A(1)$ symmetry
explicitly in the strong interactions. The simplest way to do
it\cite{Hooft} is by adding a determinant term to the Lagrangian,
\begin{equation}\label{det}
  V_{SU3SU3}= {V_{\it{U3U3}}} +\beta [\det (e^{i\theta}\Sigma)+ \det (e^{i\theta}\Sigma)^* ]\ .
\end{equation}
The addition of the complex conjugate term is required by parity,
and also by $C$ parity, since a trilinear coupling of three $C=+$
mesons must by Bose statistics be symmetric under interchange of
two mesons. We have included a $U_A(1)$ phase factor given by the
angle $\theta$. To give the pseudoscalar octet members mass (and
the $\eta ' $ a small extra mass) one conventionally adds a term $
\propto ({\rm Tr} [\Sigma M_q] +h.c.)$ where $M_q$ is a diagonal
matrix containing the chiral light quark masses.

Thereby one obtains an instructive and simple effective tree level
model for scalar and pseudoscalar mesons, essentially the $SU(3)$
version of the linear sigma model, by which one can model the
basic global symmetries of QCD and their zeroth order breaking
with the nonperturbative instanton term. Eq.(\ref{det}) is the
simplest model for the lightest mesons, which is consistent with
the symmetries of QCD. In its first formulations it has been with
us for almost 50 years\cite{LSM,UedaU3U3} and remain as a first
understanding of the symmetries involved in strong interactions.

It is the main point of this paper to show that determinant term
can give rise to three classical minima, and to show how color
symmetry enters for the lightest scalar mesons, although in an
almost hidden form.

There are  well known mathematical identities for the determinant,
which are useful for our purpose, and which we give in
eqs.(\ref{det1}-\ref{det3}) below. The first\footnote{This
identity is easily derived after diagonalisation. In terms of the
eigenvalues $a, b, c$ one has $6 \det\Sigma =6abc=(a+b+c)^3
+2(a^3+b^3+c^3)-3(a^2+b^2+c^2)(a+b+c).$} is (for $N_f=3)$,
\begin{equation}\label{det1}
  6\det \Sigma = ({\rm Tr} \Sigma )^3+2{\rm Tr}
  (\Sigma^3)-3{\rm Tr } (\Sigma^2) {\rm
Tr}(\Sigma)
   \ .
\end{equation}
In this expression each term has less symmetry ($SU(3)_F$) than
the sum \SU3SU3. In fact, each term when evaluated in terms of the
18 meson fields has many more terms than the determinant, where
most terms cancel against each other.

Another identity for a determinant\footnote{Here the matrix
element $\bar q_i q_j$ stands for the weight of the left handed
anti-quark - right handed quark component in a meson, or in a
superposition of mesons, in a rather obvious way. Thus the
dimension of $\bar qq$ is as for a meson, GeV.} $\det \Sigma_{ij}$
= $\det(\bar q_i q_j)$ comes directly from its basic definition
\begin{eqnarray}\nonumber
  \det \Sigma =\det (\bar q_i q_j ) = &\epsilon_{ijk}& \bar q_1 q_i \ \bar q_2
  q_j\ \bar q_3 q_k =\\
\label{det2}
  = \frac 1 {3!}&\delta_{ijk}^{lmn}& \bar q_lq_i\ \bar q_m
  q_j\ \bar q_n q_k  .
  \end{eqnarray}
The second  expression is written in a way which is clearly frame
inde\-pen\-dent\cite{Misner}.

Perhaps the simplest expression is obtained when the flavor sum in
eq.(\ref{det2}) is written out explicitly:
\begin{eqnarray}\nonumber
\hskip -.2cm \det \Sigma =\det (\bar q_iq_j)\hskip -.1cm =\hskip
-.3cm &+\bar u u\ \bar d d\ \bar s s-\bar u u\ \bar d s\ \bar s
d+\bar u d\ \bar d s\ \bar s u
\\ \label{det3} &-\bar u d\ \bar d u\ \bar s s+\bar u s\ \bar d u\ \bar s d-\bar u s\ \bar d d\ \bar s
u.
\end{eqnarray}
The most important physics properties of these determinant forms
are (a) The determinant is completely antisymmetric with respect
to flavor. (b) In each term one has 3 quarks and 3 anti-quarks,
and any quark flavor occurs only once, and similarly any
anti-quark flavor occurs only once. (c) It is a flavor singlet
both in the three quarks and in the three anti-quarks, and as
already noted invariant under an $SU(3)$ transformation from both
the left as well as from the right of $\Sigma$. (d) A $U_A(1)$
transformation is just a simple phase transformation
$e^{i\varphi}$ from the left and from the right, whereby only the
phase of $\Sigma$ changes by $e^{2i\varphi}$. Because of this we
have the freedom in choosing $\theta$ in eq.(\ref{det}).

It is of interest to note that in eq.(\ref{det3}) the first term
is contained only in the first term, ${(\rm Tr} \Sigma )^3$, of
eq.(\ref{det1}). The three negative terms are contained in the
third term, $-3{\rm Tr } (\Sigma^2) {\rm Tr}(\Sigma)$, of
eq.(\ref{det1}), while the two remaining positive terms in the
above equation are contained in, $2{\rm Tr}(\Sigma^3)$, of
eq.(\ref{det1}).

These equations (\ref{det}-\ref{det3}) show that the determinant
term involves a  remarkably symmetric but entangled quantum
system. In particular, note that because the three quarks or three
anti-quarks involved form a flavor singlet, any diquark subsystem
must be in the ${\bf \bar 3}_F$ representation of $SU(3)_F$. In
fact, many years ago  Jaffe\cite{Jaffe} found that in the bag
model the strongest bound diquarks are those, which are in the
antisymmetric ${\bf \bar 3}_F$ $SU(3)_F$ representation,  have
antisymmetric spin $S=0$, symmetric space (S-wave) and are
antisymmetric in color ${\bf \bar 3}_C$. Therefore he suggested a
diquark model for the lightest scalar nonet, which would have an
"inverted" mass spectrum (compared to the vector mesons), where
the $\sigma(600)$ is the lightest, followed by a $\kappa$ near 800
MeV and the $a_0(980),f_0(980)$. In fact, the model described by
eq.(\ref{det}) predicts a very broad, light sigma and the
determinant term (when including $s-d$ quark mass splitting)
shifts the $\kappa$ down from the $a_0$  by the same amount as the
K is shifted up from the $\pi$ \footnote{The contribution to
$m_\kappa^2-m^2_{a_0}=-(m^2_K-m^2_\pi) =2\beta(v_{\bar s
s}-v_{\bar d d}$)}.

The light and broad sigma\cite{NATMatts}, the $\sigma(600)$, is
now accepted as a true resonance also by the chiral perturbation
theory experts\cite{LeutSigma}. Also the expected extremely broad
$\kappa$ pole, which has been claimed in experiments\cite{E791}
has very recently\cite{Moussallam} been
determined to a remarkable accuracy 
by Roy-Steiner constraints involving crossing symmetry,
analyticity and unitarity.

 The connection between Jaffe's diquark model and
the determinant term is clear. It is natural to expect the lowest
diquarks to have spin 0 and to be in an S-wave. Since the
determinant requires any diquark to be in the ${\bf \bar 3}_F$
they must also be in the antisymmetric ${\bf \bar 3}_C$ by
spin-statistics. Thus if one wants to include color, then the
determinant term  should be multiplied by a similar factor, but
now with  color replacing flavor in the indices.

This shows the flavor-color connection through Fermi-Dirac
statistics within the scalar mesons, in a analogous way as the
color factor is needed for the proton wave function. There is,
however, one clear difference compared to Jaffe's model. The
determinant term does not describe diquark-diquark bound states
but a transition from $\bar q q$ to $\bar q  q\ \bar q  q$.
Similarly, because it describes such a transition, and not a $qqq$
state, it is not in conflict with the fact that a flavor singlet,
color singlet, S-wave, spin $\frac 1 2$ spectroscopic $qqq$ state
is forbidden by Fermi-Dirac statistics, (because of only two
degrees of freedom for spin).

Now the physical states are of course not the $\bar u u$, $\bar d
d$, $\bar s s$ appearing in eqs.(\ref{det2},\ref{det3}), but
superpositions of these (and because of the preceding discussion
also mixings with four-quark meson-meson states, with same quantum
numbers). In particular, the pure $SU(3)$ singlet states are equal
superposition of $\bar u u$, $\bar d d$, $\bar s s$. They are thus
represented by the complex matrix $\Phi=\phi\cdot {\bf 1}/\sqrt
3$, where $\phi=(s_0+ip_0)$ and where $\bf 1$ is the $3\times 3$
unit matrix. It is of some interest that these singlet terms
appear only in the first term of eqs.(\ref{det1},\ref{det3}).

 First neglect the phase angle $\theta$ in
eq.(\ref{det}). There is then a real  minimum of the potential
eq.(\ref{det}) i.e. a non zero vacuum value. (For $\mu=0$ this is
$v=\frac 1 {\sqrt 3} \phi^{min}= -\beta/(2\lambda+6\lambda ')$.)
Note that the usual positivity condition for a minimum, $v>0$,
chooses the sign of $\beta<0$ when  the sign in front of $\beta$
in eq.(\ref{pot}) is chosen positive.

But, in fact, there are three minima in the effective potential
defining 3 vacuum expectation values! Substituting $\Phi$ into
$\Sigma$ of eq.(\ref{pot}) one finds, now including the phase
$\theta$ in eq.(\ref{det}):
\begin{eqnarray}\label{pot}\nonumber
\hskip -.34cm &V(\phi)=\frac{\mu^2}{2} |\phi|^2+\frac
{\lambda+3\lambda '} 3|\phi|^4+\frac{\beta}{3\sqrt
  3}[(e^{i\theta}\phi)^3 +(e^{-i\theta}\phi^*)^3]\\
&=\frac{\mu^2}{2} |\phi|^2+\frac  {\lambda+3\lambda '}
3|\phi|^4+\frac{2\beta}{3\sqrt 3}|\phi
|^3\cos[3\theta+3\arg(\phi)] .
\end{eqnarray}
The cosine factor in the $\beta$ term makes this  potential
different from the usual "Mexican hat" potential. As an
illustrative example it is shown in fig. 1 as a contour plot in
the complex $\phi$ plane near parameter values found in
Ref.\cite{NATU3U3}.  It has
 three "hills" in the directions $\arg(\phi)=\pi-\theta$,
$\pi-\theta+2\pi/3$ and $\pi-\theta+4\pi/3$, and  three valleys in
between. Most importantly, provided $\mu^2$ is not too large and
positive, it has three minima defining three vacuum expectation
values in the downhill directions of the steepest hills

\begin{eqnarray}\label{mini}
  v_1 &=& \phi_1^{min}/\sqrt 3=v e^{-i\theta},\nonumber \\
   v_2&=& \phi_2^{min}/\sqrt 3=v e^{-i\theta+i2\pi/3}, \\
    v_3&=& \phi_3^{min}/\sqrt 3=v e^{-i\theta+i4\pi/3}\nonumber .
\end{eqnarray}

One should expect that instantons in QCD can tunnel between these
vacua and, in fact, 't Hooft\cite{Hooft} motivated the determinant
term because of instantons. The inclusion of the $\theta$ angle
shows that the three minima are all on the same footing. Although
the term (\ref{pot}) can resolve a continuous ambiguity in
$\theta$ there remains a threefold ambiguity. In the $SU(3)_F$
limit, i.e. if one neglects weak interactions and chiral quark
masses, one has the freedom to chose this chiral angle $\theta$ to
be a multiple of $2\pi/3$, such that this choice
($\theta=0,2\pi/3$ or $4\pi/3$) makes any of the three minima
 real and $>0$. Reality of
$v_ie^{i\theta}$ is required by $CP$, at least as long as weak
interactions are neglected. Expanding the meson fields around any
of these vacua $\Sigma\to\Sigma + v_i{\bf 1}$ one finds a singlet
$\eta '$ mass, $m^2_{p_0}=-6\beta |v|$=$12(\lambda+3\lambda
')|v|^2$, from the second derivative in the angular variable
($\arg(\phi)\propto p_0$) of the potential (\ref{pot}).

The scalar singlet mass is similarly obtained
$m^2_{s_0}=4(\lambda+3\lambda ')v^2$=$m^2_{p_0}/3$, or 553 MeV for
a 958 MeV $p_0$,  from the second derivative in the radial
direction $|\phi|$ of the same potential (\ref{pot}).  The scalar
octet mass is given by
$m^2_{s_{1..8}}$$=16(\lambda+3/2\lambda')v^2$=$4/3m^2_{p_0}-8\lambda
' v^2$, which means in the region of 1 GeV. The $0^{-+}$ Goldstone
octet remain in this \SU3SU3\ limit, as expected
massless

\begin{figure}
\includegraphics [width=9cm]{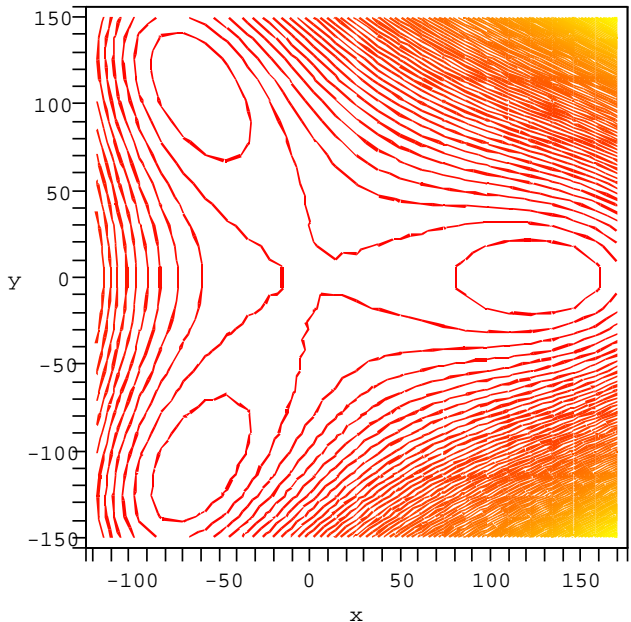}
\caption{\label{clover} An illustrative example of the potential
V($\phi$ ) of eq. (\ref{pot}) as a contour plot in the complex
$\phi$ plane. The three minima are here at $|\phi_{min}|\approx
130$ MeV. (This corresponds to an average $f_\pi$ and $f_K$ decay
constant of $130\sqrt ( 2/3) $ MeV$ \approx 106$ MeV.) The
parameters in eq.(\ref{pot}) are chosen in this illustation as
$\mu=0$, $\beta =-1700$ MeV and $\lambda+3\lambda '=11.5$. The
masses of the $SU(3)$ singlet pseudoscalar and singlet scalar
states are given by the second derivatives at any of the three
minima.}
\end{figure}
Why 3 minima? The threefold symmetry, together with $CP$, is
related to the center $Z(3)$ of the axial $SU(3)$ symmetry in
$SU_L(3)\times SU_R(3) $. Above we showed how the determinant
connects flavor and three colors because of Fermi-Dirac
statistics.  This makes three flavors special for scalar mesons,
and $N_f=3$ is also special because $SU(3)_F$ remains approximate
after symmetry breaking from the small chiral quark masses. The
symmetry breaking is small compared to $\Lambda_{QCD}$ or, here
perhaps better, compared to the $\eta '$ or  proton mass.

Thus for the meson spectrum it does not matter which of the three
$v_i$'s is chosen in the shift, $\Sigma\to\Sigma + v_i{\bf 1}$.
The meson masses remain the same since they depend on
$|v_i|^2=v^2$, but for fermions a problem appears because of the
possible phase of $v_i$. A constituent quark can get mass,
$m_q^{const}=gv_i$, through Yukawa couplings to the vacuum as in
the original linear sigma model\cite{LSM}: $g\bar q_L\Sigma
q_R+h.c. \to$ $gv_i\bar q_L q_R+h.c.$, where $g$ is a pion quark
coupling. Here $v_i$ must  be chosen chosen real for each quark. A
phase of $v_i$ like eq.(\ref{mini}) could violate parity and
charge conjugation, by which one could argue that such single free
quarks are forbidden not only by color but also $CP$.

The three minima in fig.1 are puzzling, Are these just a curiosity
of the effective model studied, or are they connected to the
longstanding strong $CP$ problem\cite{peccei} and perhaps
confinement? The axial $U_A(1)$ current is, of course, well known
not to be conserved,
because of the triangle quark graph and the gluon chiral %
anomaly. In the strong $CP$ problem one also derives from the
anomaly many different vacua connected by "large" gauge
transformations and winding numbers in the same $U_A(1)$ degree of
freedom as discussed here. The situation in our model seems
similar although the model fixes the number of vacua to three in a
cyclic fashion. Should the true vacuum be a superposition of the
three vacua as for the $\theta$ vacuum\cite{peccei}
($\sum_ne^{in\theta}|v_i>$), and should flavor symmetry be broken
in a way that maintains the permutation $Z(3)$ symmetry of the
three vacua, because of its possible connection to color?

To get a finite pseudoscalar octet mass one can, instead of a
conventional term $\propto m_q{\rm Tr}\Sigma+ h.c.$, introduce a
small term $\propto m_q({\rm Tr}\Sigma)^3+h.c.$, which retains the
$Z(3)$ symmetry (like the terms on the r.h.s. of eq.(\ref{det1})).
Similarly, instead of a conventional term $\propto{\rm Tr}(\Sigma
M_q)+h.c.$, which breaks $SU(3)_F$, one can introduce e.g.\ a term
$\propto ({\rm Tr}\Sigma)^2{\rm Tr}(\Sigma M_q)+h.c.$, which also
retains the $Z(3)$ symmetry, i.e. one still has the three equal
minima as in fig.1. These alternative forms for the symmetry
breaking results in only minor modifications for the predictions
to the mass spectrum \cite{NATtopublish}, since these depend only
on the second derivatives of the Lagrangian at the chosen minimum
$v_i$, as the singlet $\eta'$ and $\sigma$ masses in the
demonstration of Fig.1.

Fig.1 suggests that the three vacua $v_i$ and $Z(3)$ play a role
in the confinement mechanism (See Huang\cite{Huang}). Instantons
can tunnel between these minima, and the wave function of the
proton may be a superposition of states in each of the three
$v_i$. For a proton or baryon mass term $m_p\bar p_Lp_R$, where
$p$ stands for $qqq$ it would be natural that it should transform
under $U_A(1)$ like the determinant term, which also involves 3
quarks and 3 anti-quarks. Now, for a three quark system imagine a
$qqq$ wave function in the tricyclic potential of fig.1, with
three probability maxima at the three minima and let the phase of
$\bar p_Lp_R$ wind 3 times that of $\phi$, i.e. like $\det \Sigma
$. E.g. a "trial wave function" (for  $\bar p_L$ or $ p_R$) along
the chiral circle $\propto \cos (3\varphi) e^{3i\varphi }$, where
$\varphi = \arg(\phi )/2$, would do. This gives for $\bar p_Lp_R$:
$\propto [\cos (3\varphi) e^{3i\varphi }]^2 $, which transforms as
$\det\Sigma$ under a chiral rotation ($\det \Sigma \to
e^{6i\varphi}\det \Sigma  $, when $\Sigma \to e^{2i\varphi} \Sigma
$). Then for baryons, as for mesons, it does not matter which of
the three minima is chosen as real, but for a quark it would.

A perhaps better approach is to assume a quark to be a soliton,
which interpolates between two of the vacua as in the sine-Gordon
equation. A baryon is then composed of a three soliton solution
which interpolates through all three vacua starting and ending at
the same $v_i$, which is chosen  real and which remain the true
minimum after symmetry breaking.

This may open the door for a simple understanding of the
confinement mechanism. As a self-consistency check, such a
threefold rotation in the chiral angle for the baryon mass term is
consistent with the fact that baryons have integer baryon number,
but quarks have fractional baryon number of $\frac 13$. The number
three is then of topological nature as a winding number, which is
conserved although $SU(3)_F$ is broken.

In conclusion the puzzling fact that this well known effective
model has three vacua, which are illustrated in fig.~1, opens many
interesting questions. A better understanding should illuminate
the long standing strong $CP$ and confinement problems.

\section{Acknowledgements} Support  from EU RTN Contract  CT2002-0311 (Euridice) is gratefully
acknowledged.

\end{document}